\documentclass[twocolumn]{aastex62}

\usepackage{graphicx}
\usepackage[english]{babel}
\usepackage{amssymb}
\usepackage{xspace}
\usepackage{amsmath}
\usepackage{color}

\def\hi{\ifmmode {\mbox H{\scshape i}}\else H{\scshape i}\fi\xspace}
\def\hii{\ifmmode {\mbox H{\scshape ii}}\else H{\scshape ii}\fi\xspace}
\def\h2{\ifmmode {\mbox H$_2$}\else H$_2$\fi\xspace}
\def\micron{\ifmmode {\mbox $\mu$m}\else $\mu$m\fi\xspace}

\submitjournal{ApJ}

\shorttitle{A model to explain dust continuum number counts}

\shortauthors{G. Popping et al.}

\begin{document}
\title{The ALMA Spectroscopic Survey in the HUDF: A model to explain observed 1.1 and 0.85 millimeter dust continuum number counts}

\correspondingauthor{Gerg\"o Popping}
\email{gpopping@eso.org}

\author[0000-0003-1151-4659]{Gerg\"{o} Popping}
\affil{European Southern Observatory, Karl-Schwarzschild-Strasse 2, 85748, Garching, Germany}
\affil{Max Planck Institute f\"ur Astronomie, K\"onigstuhl 17, 69117 Heidelberg, Germany}

\author[0000-0003-4793-7880]{Fabian Walter}
\affil{Max Planck Institute f\"ur Astronomie, K\"onigstuhl 17, 69117 Heidelberg, Germany}
\affil{National Radio Astronomy Observatory, Pete V. Domenici Array Science Center, P.O. Box O, Socorro, NM 87801, USA}

\author{Peter Behroozi}
\affil{Department of Astronomy and Steward Observatory, University of Arizona, Tucson, AZ 85721, USA}

\author[0000-0003-3926-1411]{Jorge Gonz\'alez-L\'opez}
\affil{N\'ucleo de Astronom\'ia de la Facultad de Ingenier\'ia y Ciencias, Universidad Diego Portales, Av. Ej\'ercito Libertador 441, Santiago, Chile}
\affil{Instituto de Astrof\'{\i}sica, Facultad de F\'{\i}sica, Pontificia Universidad Cat\'olica de Chile Av. Vicu\~na Mackenna 4860, 782-0436 Macul, Santiago, Chile}

\author{Christopher C. Hayward}
\affil{Center for Computational Astrophysics, Flatiron Institute, 162 5th Ave, New York, NY 10010, USA}

\author{Rachel S. Somerville}
\affil{Center for Computational Astrophysics, Flatiron Institute, 162 5th Ave, New York, NY 10010, USA}
\affil{Department of Physics and Astronomy, Rutgers, The State University of New Jersey, 136 Frelinghuysen Rd,
Piscataway, NJ 08854, USA}

\author{Paul van der Werf}
\affil{Leiden Observatory, Leiden University, PO Box 9513, NL-2300 RA Leiden, The Netherlands}

\author{Manuel Aravena}
\affil{N\'ucleo de Astronom\'ia de la Facultad de Ingenier\'ia y Ciencias, Universidad Diego Portales, Av. Ej\'ercito Libertador 441, Santiago, Chile}

\author{Roberto J. Assef}
\affil{N\'ucleo de Astronom\'ia de la Facultad de Ingenier\'ia y Ciencias, Universidad Diego Portales, Av. Ej\'ercito Libertador 441, Santiago, Chile}

\author{Leindert Boogaard}
\affil{Leiden Observatory, Leiden University, PO Box 9513, NL-2300 RA Leiden, The Netherlands}

\author{Franz E. Bauer}
\affil{Instituto de Astrof{\'{\i}}sica and Centro de Astroingenier{\'{\i}}a, Facultad de F{\'{i}}sica, Pontificia Universidad Cat{\'{\o}}lica de Chile, Casilla 306, Santiago 22, Chile}
\affil{Millennium Institute of Astrophysics (MAS), Nuncio Monse{\~{n}}or S{\'{o}}tero Sanz 100, Providencia, Santiago, Chile}
\affil{Space Science Institute, 4750 Walnut Street, Suite 205, Boulder, Colorado 80301}

\author{Paulo C.~Cortes}
\affil{Joint ALMA Observatory - ESO, Av. Alonso de C\'ordova, 3104, Santiago, Chile}
\affil{National Radio Astronomy Observatory, 520 Edgemont Rd, Charlottesville, VA, 22903, USA} 

\author{Pierre Cox}
\affil{Institut d’astrophysique de Paris, Sorbonne Universit\'e, CNRS, UMR 7095, 98 bis bd Arago, 7014 Paris, France}

\author{Tanio D\'iaz-Santos}
\affil{N\'{u}cleo de Astronom\'{\i}a, Facultad de Ingenier\'{\i}a, Universidad Diego Portales, Av. Ej\'{e}rcito 441, Santiago, Chile}

\author[0000-0002-2662-8803]{Roberto Decarli}
\affil{INAF—Osservatorio di Astrofisica e Scienza dello Spazio, via Gobetti 93/3, I-40129, Bologna, Italy}

\author{Maximilien Franco}
\affil{AIM, CEA, CNRS, Universit\'e Paris-Saclay, Universit\'e Paris
Diderot, Sorbonne Paris Cit\'e, 91191 Gif-sur-Yvette, France}
\affil{Centre for Astrophysics Research, University of Hertfordshire, Hatfield, AL10 9AB, UK}

\author{Rob Ivison}
\affil{European Southern Observatory, Karl-Schwarzschild-Strasse 2, 85748, Garching, Germany}
\affil{Institute for Astronomy, University of Edinburgh, Royal Observatory, Blackford Hill, Edinburgh EH9 3HJ}

\author[0000-0001-9585-1462]{Dominik Riechers}
\affil{Department of Astronomy, Cornell University, Space Sciences Building, Ithaca, NY 14853, USA}
\affil{Max Planck Institute f\"ur Astronomie, K\"onigstuhl 17, 69117 Heidelberg, Germany}
\affil{Humboldt Research Fellow}

\author{Hans--Walter Rix}
\affil{Max Planck Institute f\"ur Astronomie, K\"onigstuhl 17, 69117 Heidelberg, Germany}

\author{Axel Weiss}
\affil{Max-Planck-Institut f\"ur Radioastronomie, Auf dem H\"ugel 69, 53121 Bonn, Germany}

\begin{abstract}
We present a new semi-empirical model for the dust continuum number counts of galaxies at 1.1 millimeter and 850 \micron. Our approach couples an observationally motivated model for the stellar mass and SFR distribution of galaxies with empirical scaling relations to predict the dust continuum flux density of these galaxies. Without a need to tweak the IMF,  the model
reproduces the currently available observations of the 1.1 millimeter and 850
\micron number counts, including the observed flattening  in the  1.1 millimeter number counts below 0.3 mJy \citep{Gonzalez2019numbercounts} and the number counts in discrete bins of different galaxy properties. Predictions of our work include : (1) the galaxies that dominate the number counts at flux densities below 1 mJy
(3 mJy)  at 1.1 millimeter (850 \micron) have redshifts between $z=1$ and $z=2$, stellar masses of $\sim 5\times10^{10}~\rm{M}_\odot$, and dust masses of $\sim 10^{8}~\rm{M}_\odot$; (2) the flattening in the observed 1.1 millimeter  number counts corresponds to the knee of the 1.1 millimeter luminosity function.
A similar flattening is predicted for the number counts at 850 \micron; (3) the model reproduces the redshift distribution of current 1.1 millimeter detections; (4) to efficiently detect large numbers of galaxies through their
dust continuum, future surveys should scan large areas once reaching a 1.1 millimeter flux density of 0.1 mJy rather than integrating to fainter fluxes. Our modeling framework also suggests that the amount of information on galaxy physics that can be extracted from the 1.1 millimeter and 850 \micron number counts is almost exhausted.

\end{abstract}

\keywords{galaxies: formation, galaxies: evolution, galaxies:
  high-redshift, galaxies: ISM, ISM: molecules}


\section{Introduction}
Dust-obscured star-formation contributes importantly to the cosmic
star-formation history of our Universe \citep[see the review
by][]{Madau2014}. Ever since the infrared (IR) extragalactic
background light (EBL) was first detected by the Cosmic Background
Explorer (COBE), it has become clear that the IR contributes to about
half of the total EBL \citep{Puget1996,Fixsen1998}. Understanding which galaxies are responsible
for the IR EBL, is therefore a key requirement towards understanding
which galaxies contribute most actively to the dust-obscured cosmic
star-formation thereby providing critical constraints for galaxy
formation models \citep{Granato2000,Baugh2005,Fontanot2009,Somerville2012,Cowley2015}.

A commonly used approach to better quantify the IR EBL has been to measure
the number counts of galaxies at IR wavelengths. Because of the
negative k--correction, the preferred
wavelength range to do this has been the sub-millimeter and millimeter
regime. The first efforts to measure number counts were carried out
with single dish instruments such as SCUBA and LABOCA
(\citealt{Eales2000,Smail2002,Coppin2006,Knudsen2008,Weiss2009} and see
\citealt{Casey2014} for a more extensive review). These efforts have
been paramount for our understanding of the IR EBL, but typically
suffered from a lack of sensitivity and from source blending due to poor
angular resolution.

The advent of the Atacama Large Millimeter/sub-millimeter Array (ALMA)
has opened up a new means to quantify the IR EBL. In particular, the
superior sensitivity of ALMA allows for a better quantification of the IR EBL
down to fainter limits. This is further aided by a higher angular
resolution that can overcome source blending. Indeed, since ALMA started
operating a large number of works in the literature have contributed
to better quantifying millimeter and sub-millimeter number counts
\citep{Hatsukade2013,Ono2014,Carniani2015,Oteo2016,Dunlop2016,Aravena2016,Hatsukade2016,Fujimoto2016,Umehata2017,Franco2018,
  Munoz2018, Gonzalez2019numbercounts}. \citet{Aravena2016}, \citet{Fujimoto2016}, and \citet{Munoz2018}  have pushed the quantification of 1.2 millimeter number counts
down to flux densities of 0.3 and 0.02 mJy, respectively. 
\citet{Fujimoto2016} reached this
conclusion by taking advantage of lensing through a cluster.  More recently, \citet{Munoz2018} also measured the number counts 
of galaxies at 1.1 millimeter down to 0.01 mJy taking advantage of
lensing. Although focusing on lensed sources has proven to be an
efficient way to reach faint flux densities,  uncertainties in the lensing model
complicate the precise derivation of the faint number
counts. \citet{Aravena2016} on the other hand reached flux densities
of 0.3 mJy as a part of the
ASPECS pilot project \citep{Walter2016}, targeting the 1.2 mm emission
in a contiguous blank region on the sky corresponding to $\sim1$ arcmin$^2$.

\citet{Gonzalez2019numbercounts} present the deepest 1.2 mm
continuum images obtained to date in a contiguous area over the sky
(4.2 arcmin$^2$), reaching number count statistics down to an rms flux density
of 9.5$\mu\rm{Jy}$ per beam. This work was based on the band
6 component of the full ASPECS survey,
whose first results were presented in \citet{Aravena2019},
\citet{Boogaard2019}, \citet{Decarli2019},\citet{Gonzalez2019},  and
\citet{Popping2019}. \citet{Gonzalez2019numbercounts} found that the 1.2 mm number counts flatten below
flux densities of $\sim 0.3$ mJy. These results are similar to the earlier findings at less
significance by
\citet{Munoz2018} based on lensed sub--mm emission in three galaxy
clusters. \citet{Gonzalez2019numbercounts} was furthermore able to decompose the 1.2 millimeter number counts in bins of different galaxy properties (redshift, stellar mass, star formation rate, and dust mass). Now that the shape and normalization of the 1.2 mm number counts are well characterised by ALMA, as well as how these decompose in bins of different galaxy properties, it is crucial to put these observations in a theoretical framework. 

In this paper we present a new semi--empirical approach to model the
1.1 millimeter and 850 $\mu$m number counts of galaxies. This model is
designed to explore how the number counts are built up by
contributions from galaxy samples at different redshifts and varying
galaxy properties (i.e., the star formation rate (SFR), stellar mass, and
dust mass). In particular, we aim to address the cause for the flattening in the 1.2 millimeter number counts of
galaxies, and if a similar flattening is to be expected in the 850
$\mu$m number counts. To this aim, we explore which galaxies are
responsible for different parts of the (sub-)millimeter number counts
of galaxies. Based on our findings, we furthermore discuss the best
strategies to detect large numbers of galaxies through their dust continuum.

The paper is outlined as follows. We present the model in Section
\ref{sec:model}. We present the
predictions by the model and how they compare to and explain the
observational data in Section \ref{sec:results}. We discuss our
findings in Section \ref{sec:discussion} and summarise them and draw
conclusions in Section \ref{sec:conclusion}.  Throughout this paper we
adopt a flat $\Lambda$CDM cosmology, with parameters ($\Omega_M =
0.307$, $\Omega_\Lambda = 0.693$, $h=0.678$, $\sigma_8 = 0.823$, and
$n_s = 0.96$) similar to Planck 2018 constraints
\citep{Planck2018}. We furthermore adopt a \citet{Chabrier2003} stellar initial mass function.

\section{Model description}
\label{sec:model}
This section describes our methodology to predict the sub--mm continuum
flux density of galaxies. In summary, we start with mock light cones (i.e.,
a continuous model galaxy distribution from $z=0$ to $z=10$ over an area on
the sky) created
by the UniverseMachine \citep{Behroozi2018}, which assigns galaxy
properties (stellar mass, SFR) to haloes based on observationally
constrained relations. We then use a number of empirical relations to
assign dust masses to each galaxy. We calculate the 850 $\mu$m and
1.1 millimeter flux density of galaxies following the fits presented
in \citet{Hayward2011} and \citet{Hayward2013SHAM} as a function of galaxy SFR and dust mass.

\subsection{Generating mock lightcones}
The \textsc{UniverseMachine} is an empirical model of galaxy formation
that infers how the star formation rates of galaxies depend on host halo
mass, halo mass accretion rate, and redshift via forward modeling
\citep{Behroozi2018}.  Given a guess for the SFR--halo relationship,
the \textsc{UniverseMachine} applies the relationship to a dark matter
halo catalog and generates an entire mock universe.  This mock
universe is observed in the same way as the real Universe, and galaxy
statistics (including stellar mass functions, specific star formation
rates, galaxy clustering, luminosity functions, and quenched
fractions, among others) are compared to evaluate the likelihood for
the given SFR--halo relationship to be correct.  This likelihood is
then fed to a Markov Chain Monte Carlo algorithm that explores the
posterior distribution of SFR--halo relationships that match
observations.  The model was compared to galaxy observations from
among others the
SDSS, PRIMUS, CANDELS, zFOURGE, and ULTRAVISTA surveys over the range
$z=0$ to $z=10$; for full details of the modeling and data, see
\cite{Behroozi2018}.  The underlying dark matter simulation was
\textit{Bolshoi-Planck}, which resolves halos down to $10^{10} \rm{M}_\odot$ (hosting galaxies down to $10^7  \rm{M}_\odot$) in a periodic cosmological region that is 250 Mpc h$^{-1}$ on a side \citep{Klypin14,RP16b}.  Halo finding and merger tree construction were performed by the \textsc{Rockstar} and \textsc{Consistent-Trees} codes, respectively \citep{Rockstar,BehrooziTree}.

The lightcones used in this paper are based on the bestfit \textsc{UniverseMachine} DR1 SFR--halo relationship.  This relationship was used to generate a mock catalog containing galaxy stellar masses and star formation rates for every halo (and subhalo) in \textit{Bolshoi-Planck} at every redshift output (180, equally spaced in $\log(a)$ from $z\sim 20$ to $z=0$).  Eight lightcones were generated for the CANDELS GOODS-S field footprints by choosing random locations within the simulation volume and then selecting halos along a random line of sight, tiling the periodic simulation volume as necessary.  When selecting halos, the cosmological distance along the lightcone was used to determine the closest simulation redshift output to use.  The final lightcones include galaxy stellar masses, star formation rates, sky positions, and redshifts (including both cosmological redshift and redshift due to peculiar velocities), as well as full dark matter halo properties.

\subsection{Assigning (sub-)mm luminosities to galaxies}
\citet{Hayward2011} and \citet{Hayward2013hydro} presented fitting functions for the (sub-mm) flux
densities of galaxies based on their SFR and dust mass. These fitting
functions were derived by running the \textsc{SUNRISE}
\citep{Jonsson2006} dust radiative transfer code on smoothed
particle hydrodynamics simulations of isolated and merging
galaxies. The authors found that the 850 $\mu$m and 1.1 millimeter
flux density of IR--bright galaxies (down to 0.5 mJy) can be well described by
\begin{equation}
\label{eq:submm850}
S_{850~\mu\rm{m}} = 0.81 \rm{mJy}\left
  (\frac{\rm{SFR_{\rm obscured}}}{100~\rm{M}_\odot\rm{yr}^{-1}}\right )^{0.43}\left(\frac{M_d}{10^8~\rm{M}_\odot}\right)^{0.54},
\end{equation}
and 
\begin{equation}
\label{eq:submm1.1}
S_{1.1~\rm{mm}} = 0.35 \rm{mJy}\left
  (\frac{\rm{SFR_{\rm obscured}}}{100~\rm{M}_\odot\rm{yr}^{-1}}\right )^{0.41}\left(\frac{M_d}{10^8~\rm{M}_\odot}\right)^{0.56},
\end{equation}
where $S_{850~\mu\rm{m}}$ and $S_{1.1~\rm{mm}}$ mark the 850 $\mu$m and
1.1 millimeter flux density, and $\rm{SFR_{\rm obscured}}$ and $M_d$ the dust obscured
SFR of galaxies and dust mass of a galaxy,
respectively. \citet{Hayward2011} find that these functions recover
the sub-mm flux (brighter than 0.5 mJy) at these wavelengths of simulated galaxies to within a scatter of 0.13 dex
in the redshift range $z\sim 1$--6 (we include this scatter when we
calculate fluxes). The apparent redshift independence of this
relation is a natural result of the negative k--correction in the
millimeter range of the galaxy spectral energy distribution. This fit
under predicts the flux of galaxies significantly at $z<0.5$. Because
of the change in normalization of the main-sequence of star-formation from
$z=0.5$ to $z=0$ \citep[e.g.,][]{Speagle2014} we do not expect these
galaxies to contribute significantly to the total sub--mm flux density (as we will see in Sec. \ref{sec:which_galaxies}). 
Furthermore, the volume probed by a survey  in the
redshift range $z=0$--0.5 is only a small fraction of the total volume
from $z=0$ to $z=8$.\footnote{Our results regarding the flattening of the number counts are not sensitive to the uncertainties in the estimated flux within the $z=$0--0.5 redshift range. Even in the extreme scenario that the predicted fluxes at $z<1$ are too low by an order of magnitude do we still recover the flattening in the number counts (see also the redshift distribution of the number counts in Figure \ref{fig:number_counts_redshift}).} We furthermore do not include a correction for the cosmic microwave background (CMB) as a background radiation field in this work. Our methodology does not provide the actual dust temperature of the simulated galaxies, from which a correction factor can be estimated following \citet{dacunha2013}. If we assume a dust temperature of 20 Kelvin, we expect that 90\% of the intrinsic flux emitted by galaxies at $z=3$ is observed against the CMB background. There have been works suggesting the dust temperature of galaxies evolves to even higher temperatures (40 Kelvin and above at $z>3$) as a function of lookback time \citep[e.g.,][]{Bouwens2016, Narayanan2018}. At these temperatures more than 95\% of the intrinsic flux is observed against the CMB background at $z<5$. We are therefore confident that (at least for the regime where we can directly compare our model to observations) the CMB won't alter our results significantly.

The dust obscured SFR can be described as 
\begin{equation}
\rm{SFR}_{\rm obscured} =
f_{\rm obscured}\rm{SFR}_{\rm total},
\end{equation}
 where $f_{\rm obscured}$
corresponds to the obscured fraction of star formation and
$\rm{SFR}_{\rm total}$ corresponds to the total SFR of galaxies (the
sum of the obscured and unobscured fraction). To calculate $f_{\rm
  obscured}$ we use the empirical relation derived by
\citet{Whitaker2017} between the obscured fraction of star formation
and the stellar mass for main-sequence galaxies in the redshift range from $z=0.5$
to $z=2.5$. We assume that this empirical fit extends towards higher redshift and also applies for galaxies above the main-sequence. \citet{Hayward2013hydro} do not make an explicit distinction between unobscured and obscured star formation in their fitting functions (i.e., they implicitly assume that all star formation is dust obscured). To quantify the effect of introducing the parametrization by \citet{Whitaker2017} we explore the scenario where $f_{\rm obscured}$ is set to one in Appendix A. We find that the predicted number counts are almost identical to the predictions by our fiducial.

To calculate the dust mass $M_d$ of galaxies, we use a strategy similar
to the one presented in \citet{Hayward2013SHAM}. We first calculate the
total gas mass of galaxies as described in \citet{Popping2015SHAM}. The authors determine gas masses for galaxy catalogues
generated using sub-halo abundance matching models. In summary, the authors calculate
what gas mass a galaxy must have to have a SFR equal to the SFR
obtained from the sub-halo
abundance matching model. This is done by randomly picking a gas mass
for a galaxy and assuming that the gas and
stellar mass
of this galaxy are distributed exponentially, with a scale length given by
the stellar mass -- size relation of galaxies as found by
\citet{vanderWel2014}. At every point in the disc, the gas is then divided into a
molecular and an atomic component, following the empirical relation
determined by \citet{Blitz2006} which relates the mid-plane pressure
acting on the gas disc to the molecular hydrogen fraction. The SFR
surface density is then calculated as a function of the
molecular hydrogen surface density following \citet{Bigiel2008}, but
allowing for an increased star-formation efficiency in high surface
density environments. The total SFR of a galaxy is calculated by
integrating over the entire disc. The `true' gas mass of a galaxy is
determined by iterating over gas masses till the SFR calculated
following these empirical relations equals the SFR provided by the
sub-halo abundance matching model. A more detailed description of this
method is given in \citet{Popping2015SHAM} and \citet{Popping2015CANDELS}.

Once the total cold gas mass of a galaxy is known, we estimate the dust
mass of this galaxy by multiplying it with a dust--to--gas ratio. We
use the fit presented in \citet{DeVis2019} between dust--to--gas ratio and gas-phase metallicity
of galaxies of local galaxies to estimate a dust--to--gas ratio. Theoretical simulations have suggested that the relation
between dust--to--gas ratio and gas-phase metallicity hardly evolves
between redshifts $z=0$ and $z=6$ 
(e.g., \citealt{Feldmann2015}, \citealt{Popping2017dust}, though see
\citealt{Hou2019} who suggest that the normalization of the relation
between dust--to--gas ratio and gas--phase metallicity decreases at
$z>3$). The gas-phase metallicity of galaxies is estimated as a
function of the stellar mass and redshift by fitting the results presented
in \citet[see also \citealt{Zahid2014}]{Zahid2013}. The metallicities are converted to the
same metallicity calibration as used in \citet{DeVis2019} following
the approach presented in \citet{Kewley2008}. \citet{Zahid2013} presents
metallicities for a sample of galaxies out to $z\sim 2.26$ and we assume that the
redshift dependent fit to the mass-metallicity relation extends
towards higher redshifts. A similar approach was also adopted by
\citet{Imara2018} to assign dust masses to galaxies based on empirical
scaling relations.

Throughout this process we use the stellar mass and SFR predicted by
the \textsc{UniverseMachine} as input for the empirical relations.  To
 account for the fact that empirical relations are based on
 observationally derived stellar masses and SFRs and not on the
 intrinsic stellar mass and SFR of a galaxy, we make use of the predictions for
 galaxy properties from the \textsc{UniverseMachine} that account for
 observational effects and errors. Each of the adopted empirical
 relations has an intrinsic error associated to it. To
 account for this, we run 100 realizations of the model, sampling over
 errors in the empirical relations. In the Appendix of this paper we 
  explore alternative empirical relations with the aim of developing a sense of how robust our results are against
  our assumptions. We do not account for
  blending effects and gravitational lensing when modeling  number
  counts as our analysis focuses on flux densities for which blending
  is not thought to significantly contribute to the number counts \citep[e.g.,][]{Hayward2013SHAM}.

To test the validity of our model we compare the 1.1 millimeter flux predicted for the galaxies observed in \citet{Gonzalez2019numbercounts} based on their observed stellar mass, SFR, and redshift to the observed fluxes. We find that the mean ratio between the predicted and observed 1.1 millimeter flux densities for these objects is 1.05, with a standard deviation of 0.81.

\begin{figure*}
\centering
\includegraphics[width = \hsize]{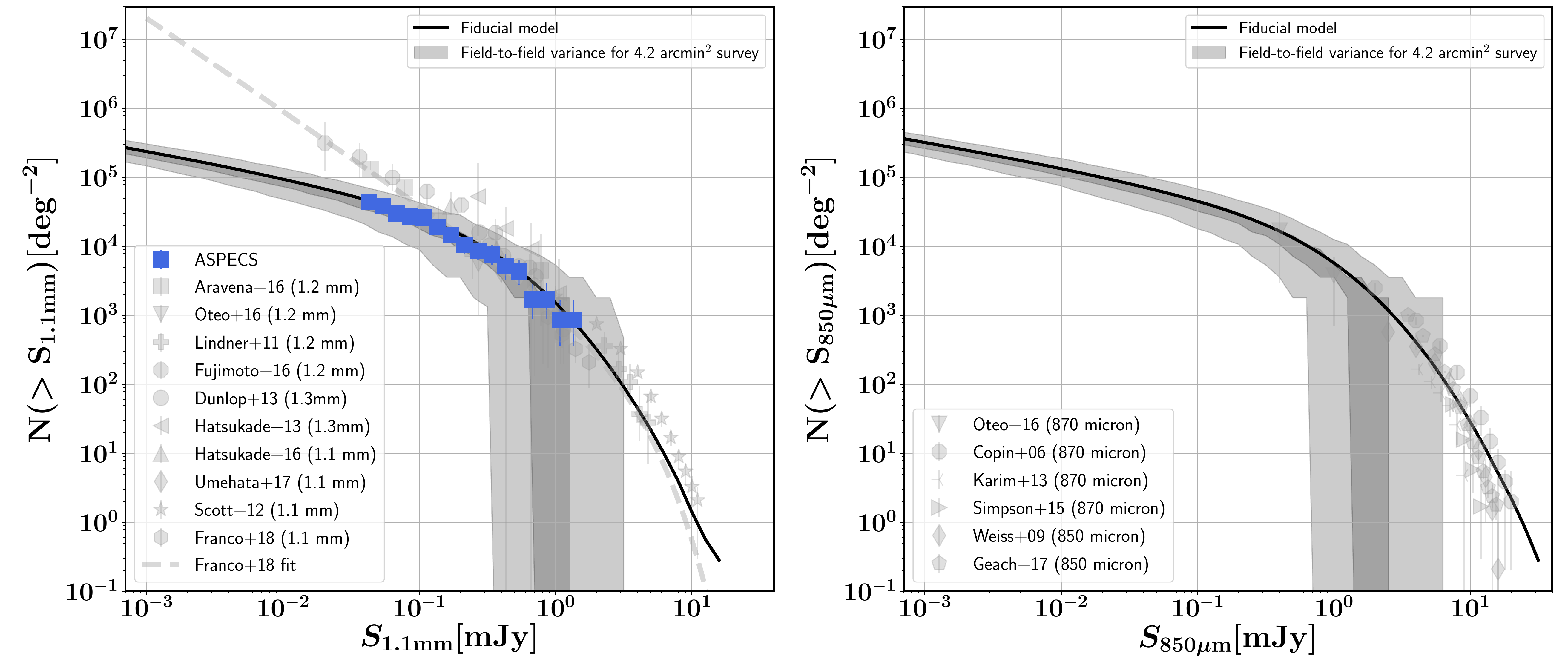}
\caption{The 1.1 millimeter (left) and 850 \micron (right) galaxy number counts. The black solid
  lines mark our predictions for the number counts when accounting for
  all the galaxies in
  the entire simulated lightcone. The dark-- and light--gray shaded areas mark the one-- and
  two--sigma scatter due to
  field--to--field variance, assuming a survey with the size of ASPECS
  (i.e., 4.2 arcmin$^2$). The model predictions are compared to a
  literature compilation of number counts, where the dashed line corresponds to the Schechter fit presented by Franco et al. to their literature compilation. The blue points show the. number counts derived from ASPECS \citep{Gonzalez2019numbercounts}
\label{fig:number_variance}}
\end{figure*}

\begin{figure}
\centering
\includegraphics[width = \hsize]{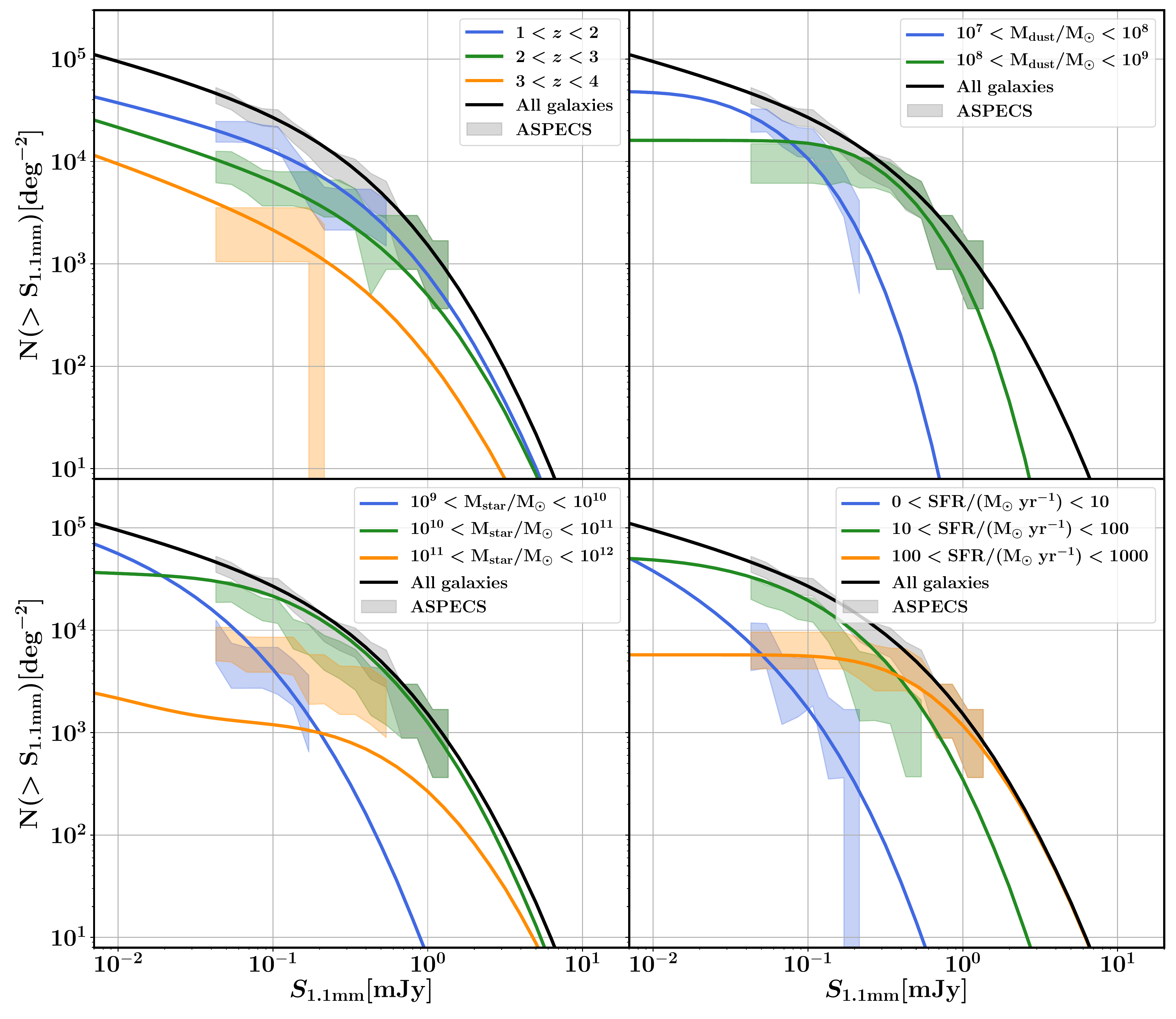}
\caption{The predicted and observed 1.1 millimeter galaxy number counts in bins of redshift (top left), dust mass (top right), stellar mass (bottom left), and SFR (bottom right). The solid lines correspond to the model predictions, whereas the shaded areas show the ASPECS observations. 
\label{fig:number_counts_JorgeCompare}}
\end{figure}

\begin{figure*}
\centering
\includegraphics[width = \hsize]{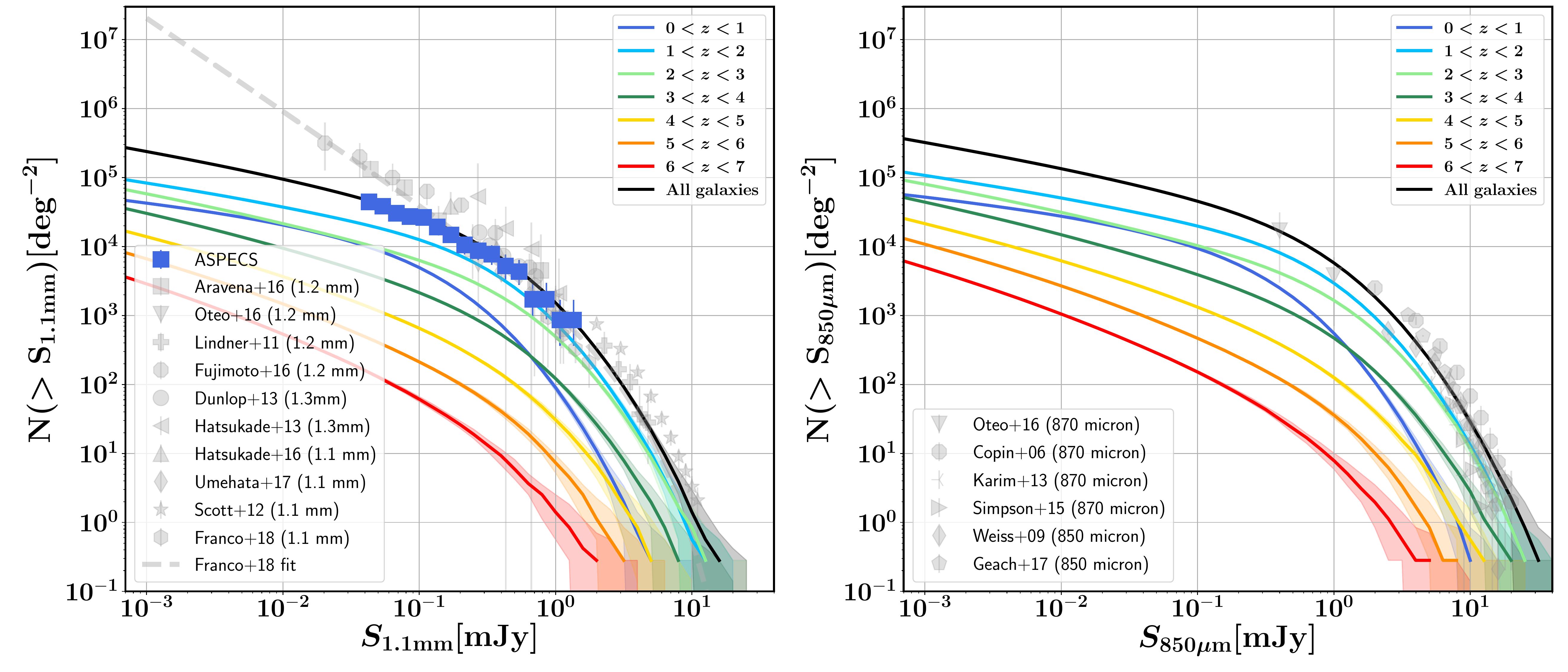}
\caption{The 1.1 millimeter (left) and 850 \micron (right) galaxy number counts. The black solid
  lines mark our predictions for the number counts when accounting for
  all the galaxies in
  the lightcone  (as shown in Fig.~\ref{fig:number_variance}). The coloured lines mark the number counts when
  selecting galaxies based on their redshift. The color shading
  corresponds to the two-sigma scatter when sampling over the
  intrinsic scatter of the empirical scaling relations. The model predictions are compared to a
  literature compilation of number counts as in Fig.~\ref{fig:number_variance}. The 1.1 mm number counts are dominated by galaxies at
  $z=$1--2, with additional contributions from galaxies up to $z=3$ at
  the brightest fluxes and galaxies in the range $z=$0--1 at the
  faintest fluxes. 
\label{fig:number_counts_redshift}}
\end{figure*}

\begin{figure*}
\centering
\includegraphics[width = 0.9\hsize]{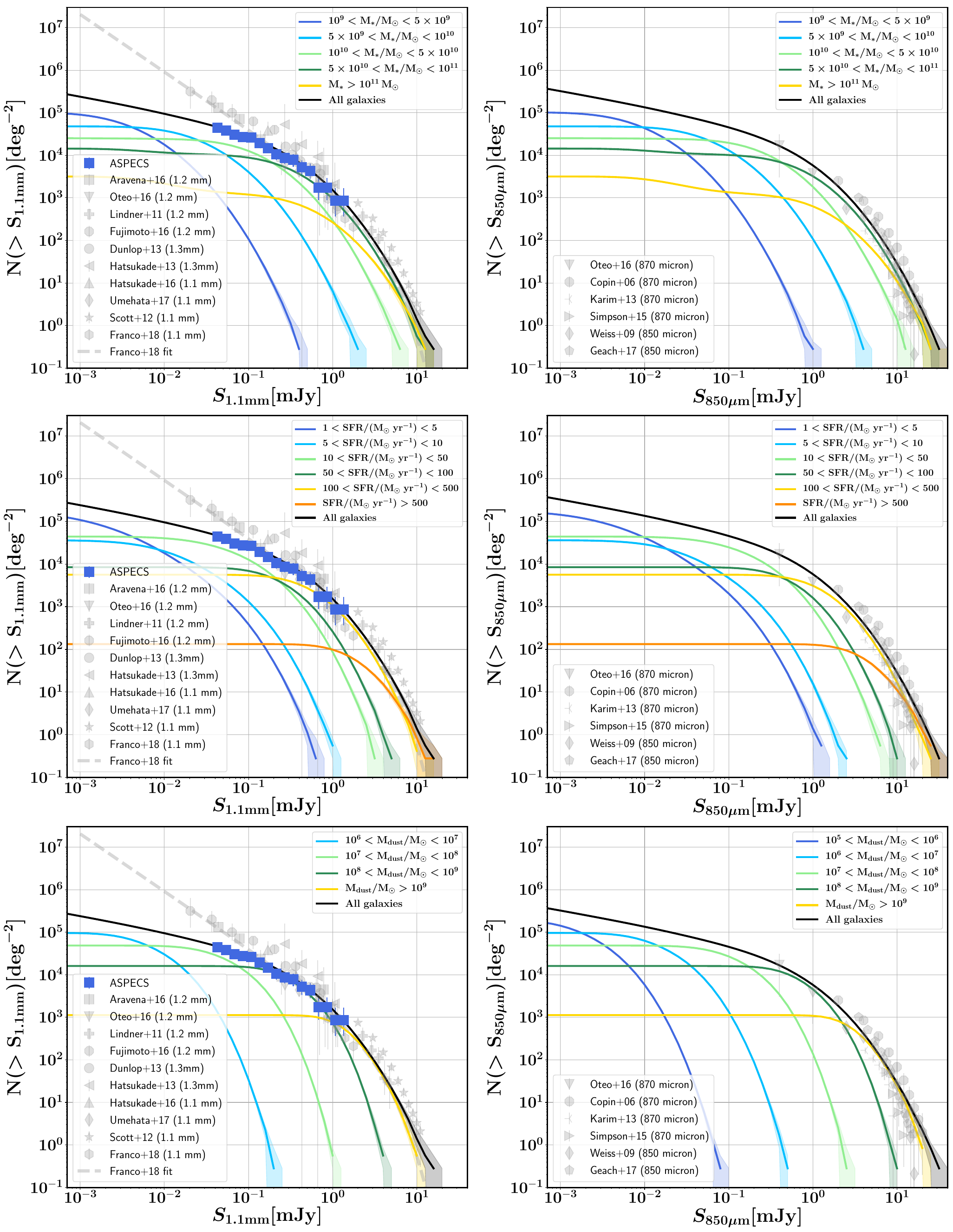}
\caption{The 1.1 millimeter (left) and 850 \micron (right) galaxy number counts of galaxies, broken up by different galaxy properties (integrated over all redshifts). The black solid
  lines mark our predictions for the number counts when accounting for
  all the galaxies in
  the lightcone (as shown in Fig.~\ref{fig:number_variance}). The coloured lines mark the number counts when
  selecting galaxies based on their stellar mass (top row), SFR (middle row), and dust mass (bottom row). The color shading
  corresponds to the two-sigma scatter when sampling over the
  intrinsic scatter of the empirical scaling relations. The model predictions are compared to a
  literature compilation of number counts as in Fig.~\ref{fig:number_variance}. 
\label{fig:number_counts_properties}}
\end{figure*}

\section{Results}
\label{sec:results}
In this Section we present our predictions for the 1.1 millimeter and
850 \micron number counts of galaxies, specifically focusing on how
they compare to current observations and which galaxies are
responsible for the number counts at different flux
densities. Throughout this paper we compare our model predictions to a
set of observations taken from \citet{Coppin2006}, \citet{Weiss2009}, \citet{Lindner2011}, \citet{Scott2012},
\citet{Hatsukade2013}, \citet{Karim2013}, \citet{Simpson2015},
\citet{Aravena2016}, \citet{Dunlop2016}, \citet{Fujimoto2016},
\citet{Hatsukade2016}, \citet{Oteo2016}, \citet{Umehata2017},
\citet{Geach2017}, \citet{Franco2018}, and
\citet[the deepest survey at 1.2 millimeter over a contiguous area
on the sky to date]{Gonzalez2019numbercounts}. This compilation
includes observations based on single-dish instruments as well as with ALMA.
These observations were carried out over a range of wavelengths, and
scaled to 1.1 millimeter and 850 \micron fluxes such that $S_{\rm 1.1mm}/S_{\rm 1.2mm} =
1.36$, $S_{\rm 1.1mm}/S_{\rm 1.3mm} = 1.79$, and $S_{\rm 870\mu\rm{m}}/S_{{\rm
    850}\mu\rm{m}} = 0.92$, assuming a dust emissivity index $\beta = 1.5-2.0$ \citep[e.g.,][]{Draine2011} and a temperature of 25--40 Kelvin (e.g.,\citealt{Magdis2012, Schreiber2018}). We first present the model number counts
and how field--to--field variance affects the derived number counts. We
then break up the number counts in bins of redshift, dust mass,
stellar mass, and SFR. We finish by showing the redshift distribution
of galaxies compared to observations.

\subsection{The (sub-)mm number counts of galaxies and field-to-field variance}
We present the 1.1 millimeter and 850 \micron flux density number
counts of galaxies in Figure \ref{fig:number_variance} (black solid lines). The number counts predicted by the model are in good
agreement with the ASPECS data, both at 1.1 millimeter and at 850 \micron
over the full flux density range where observations are available.  
We predict a flattening in the
number counts of galaxies for flux densities below $\sim 0.3$ mJy at 
1.1 millimeter, similar to the flattening
found by \citet{Gonzalez2019numbercounts}. We also find a flattening in
the 850 \micron number counts around a flux density of $\sim
1$ mJy. The predicted number counts lie below the observations by \citet{Fujimoto2016}, who derived their number counts based on uncertain lensing models. \citet{Aravena2016} calculated their number counts based on a significantly smaller area and simpler analysis techniques. A more detailed description of the source of the discrepancy is given in \citep{Gonzalez2019numbercounts}. 

Since one of the specific aims of this paper is to assess the origin of the
flattening in the 1.1 mm number counts detected by
\citet{Gonzalez2019numbercounts}, we show the number counts derived for
the entire simulated area, as well as the number counts derived for a
simulated area corresponding to the ASPECS survey. To this aim we
calculate the number counts in 100 randomly drawn sub-areas covering 4.2 arcmin$^2$ (the area covered by ASPECS) on the
sky. The number counts of the full simulated volume are depicted as a
black solid line, whereas the one- and two-sigma scatter when
calculating the number counts in the areas corresponding to ASPECS are depicted as gray shaded regions. There are two noteworthy results with regards to cosmic variance. First of all, at flux densities fainter
than 1 (3) mJy when focusing on 1.1 millimeter (850 \micron) emission,
the typical two-sigma 
scatter due to field-to-field variance is only a factor of 1.5 and the flattening in the number counts is always recovered. Second, due to the
small area covered, sources brighter than 1 mJy (at 1.1 millimeter, 3
mJy at 850 \micron) are typically missed by surveys targeting only 4.2
arcmin$^2$ on the sky (see also Figure
\ref{fig:redshift_distribution}).

\subsection{Which galaxies are the main contributors to  the number counts?}
\label{sec:which_galaxies}
The depth of the ASPECS survey combined with the rich ancillary data available in the HUDF allowed \citet{Gonzalez2019numbercounts} to decompose the observed 1.2 millimeter number counts in bins of stellar mass, dust mass, SFR, and redshift. We compare our model predictions to these observations in Figure \ref{fig:number_counts_JorgeCompare}. We find decent agreement between the observations and model predictions when breaking up the number counts in bins of redshift, dust mass, and SFR. When breaking up the number counts in bins of stellar mass, we find that the contribution of galaxies with stellar masses between $10^9$ and $10^{10}\,\rm{M}_\odot$ is well reproduced. Our model predicts a contribution to the number counts below 0.5 mJy by galaxies with a stellar mass between $10^{10}$ and $10^{11}$ solar masses that is too large (up to a factor of two). The predicted contribution by galaxies with larger stellar masses in this flux density range is too small (up to a factor of three) compared to the observations. Tests have shown that when we change the stellar mass bins (e.g., from $10^{10.5}$ to $10^{11.5}\,\rm{M}_\odot$) the agreement between models and observations is much better. This suggests that the discrepancy is (at least partially) driven by uncertainties in the observed stellar masses that can easily be of the order 0.3 dex \citep{Leja2018}. We have furthermore not taken the effects of cosmic variance into account in this comparison, which can be non-negligible for the bins with highest stellar masses (\citealt{Moster2011}, since the ASPECS survey only covers an area of 4.2 squared arcsec in ALMA band 6). The good agreement between the model predictions is encouraging and opens up the opportunity to explore the model further to better understand which galaxies contribute to the number counts at different flux densities.

We show the number counts of galaxies in different redshift bins in
Figure \ref{fig:number_counts_redshift}. Galaxies at $z>3$ make up for a small fraction of the total number counts at 1.1 millimeter
and 850 \micron. The number counts
are made up by an equal contribution of galaxies in the redshift range
$z=$2--3 and $z=$1--2 for flux densities brighter than  $\sim$3 ($\sim 6$)
mJy at 1.1 millimeter (850 \micron). At lower flux densities, the largest
contribution to the number counts comes from galaxies in the redshift
bin $z=$1--2. Galaxies at $z<1$ hardly contribute to the number
counts at flux densities larger than $\sim 0.1$ mJy at both
wavelengths, whereas they contribute more importantly to the number counts at
fainter fluxes (although still a factor of 2 less than galaxies at
$z=$1--2). There is a clear flattening visible in the number counts of
galaxies at all redshifts. The galaxy population that contributes most to the total (all redshifts) number counts at flux densities of 0.3 mJy at 1.1 millimeter (1 mJy at 850 \micron, this corresponds to the flux density below which the total number counts rapidly flatten  ) consists of galaxies with redshifts in the range $z=$1--2.

In Figure \ref{fig:number_counts_properties} we show the number counts of
galaxies in bins of stellar mass. As the flux density
increases the number counts are dominated by more massive
galaxies. This is a natural consequence of an increase in dust mass
and SFR of galaxies as a function of stellar mass.  Galaxies with stellar masses around $5 \times 10^{10}\,\rm{M}_\odot$ contribute most dominantly to the number counts at the flux density below which the number counts flatten (0.3 and 1 mJy at 1.1 millimeter and 850 \micron, respectively).

We show the number counts of galaxies in bins of SFR in the middle row of Figure
\ref{fig:number_counts_properties}.  Not surprisingly, we find that the
number counts at the brightest flux densities probed by observations are dominated by the
most actively star-forming galaxies (i.e., SFR
$>100~\rm{M}_\odot~\rm{yr}^{-1}$). Interestingly, at $\sim$0.25 (0.6)
mJy the 1.1 millimeter (850 \micron) number counts are driven by an
equal contribution from galaxies with a SFR in the bin between 10--50,
50--100, and 100--500 $\rm{M}_\odot~\rm{yr}^{-1}$. This pivoting point
also roughly marks the location of the flattening in the number counts. At lower flux densities (but
brighter than 0.05 and 0.1 mJy for the 1.1 millimeter and 850 \micron
number counts, respectively) the number
densities are dominated by galaxies with a SFR $=$10--50
$\rm{M}_\odot~\rm{yr}^{-1}$. At even lower flux densities galaxies
with SFRs between 1 and 5 $\rm{M}_\odot\,\rm{yr}^{-1}$ are predominantly
responsible for the number counts. In the previous figures we noticed that as the flux density
increases the number counts are dominated by more massive
galaxies. Such a behavior is not seen for the SFR of galaxies. Some
bins in SFR (e.g., 5--10 and 50--100 $\rm{M}_\odot~\rm{yr}^{-1}$) are never the dominant population of galaxies responsible for the
  observed total number counts. This is because the 1.1 millimeter and
  850 \micron fluxes of galaxies depend more strongly on dust mass
  than on SFR (see Equations \ref{eq:submm1.1} and
  \ref{eq:submm850}).
  
The contribution by galaxies with different dust masses to the 1.1
millimeter and 850 \micron number counts is also presented in Figure
\ref{fig:number_counts_properties} (bottom row). Similar to the stellar mass, we find
that as the flux density increases, the number counts are dominated by
galaxies with increasing dust masses. We find that galaxies with dust masses in
the range between $10^8$ and $10^9\,\rm{M}_\odot$ contribute most strongly to the number counts at 0.3 (1.0) mJy at 1.1 millimeter (850 \micron), the flux density below which the number counts flatten.

\begin{figure*}
\centering
\includegraphics[width = \hsize]{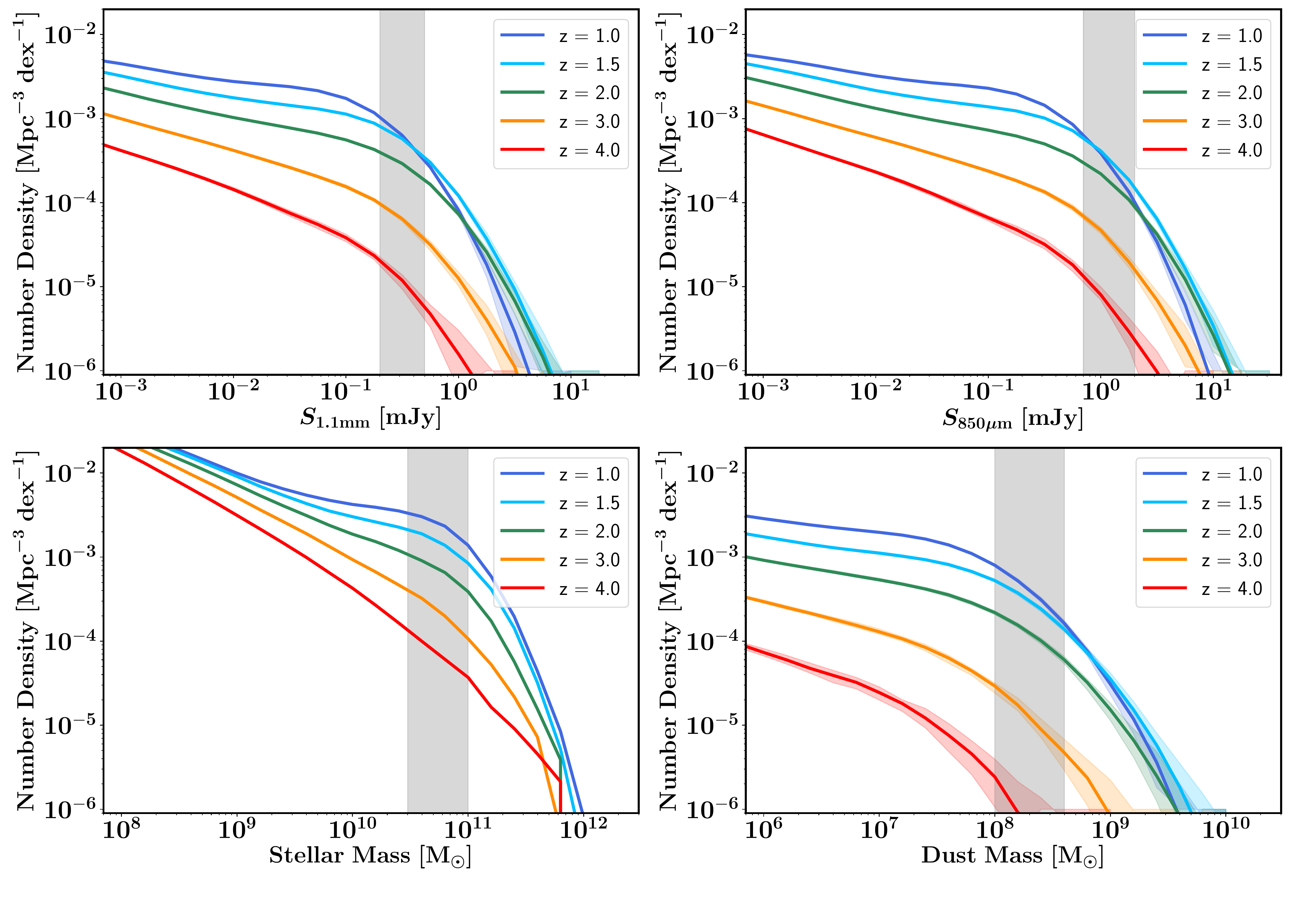}
\caption{The 1.1 millimeter luminosity function (top left),
  the 850 \micron luminosity function (top right), the
  stellar mass function (bottom left), and the dust mass function
  (bottom right) of galaxies at different redshifts. The color shading
  corresponds to the two-sigma scatter when sampling over the
  intrinsic scatter of the empirical scaling relations. The grey shaded
  band in each panel corresponds to the galaxies that contribute most
  dominantly to flux density at which the predicted flattening starts in the 1.1 millimeter and 850
  \micron number counts. The grey bands overlap with the knee of the
  respective mass/luminosity functions, suggesting that the flattening
  in number counts is a reflection of the  1.1 millimeter and 850
  \micron luminosity functions. We do not show the luminosity and mass functions at $z<1$ since the predicted flux densities at these redshifts are not reliable.
\label{fig:mass_functions}}
\end{figure*}

\subsection{The flattening in number counts corresponds to the knee
  and shallow faint end slope of the dust continuum luminosity functions}
\label{sec:mass_functions}
In the previous subsection we have seen that our model and the observations suggest that galaxies at $z=$1--2 contribute most to the flux densities at which the 1.1 millimeter and 850 \micron number counts flatten (Figure \ref{fig:number_counts_redshift}). We
have furthermore seen that the galaxies responsible for the flattening
have stellar masses around $5 \times 10^{10}~\rm{M}_\odot$, dust
masses between $10^8$ and $10^9~\rm{M}_\odot$, and SFRs in the range
between 10 and 500 $M_\sun$ yr$^{-1}$. At $z=$1--2, a stellar mass
of $5 \times 10^{10}~\rm{M}_\odot$  roughly corresponds to the stellar
mass at the knee of the stellar mass function at these redshifts \citep[e.g.,][]{Tomczak2014}. This
suggests that the flattening in the number counts is driven by the
shape of the 1.1 millimeter and 850 \micron luminosity
function at $z=$1--2 and that the flattening may actually simply reflect observations
 probing galaxies below the knee of this function. 

To test our hypothesis we switch from number counts (projected densities on the sky) to volume densities. In Figure \ref{fig:mass_functions} we show the  luminosity function (number of sources per volume element) predicted from our model as a function of redshift (cosmic time).\footnote{These are actually 1.1 millimeter and 850 \micron flux density distribution functions, but for simplicity we call them luminosity functions.} We also show the stellar mass function and dust mass
functions. We highlight the flux density and stellar (dust) mass
regime at which the flattening occurs with a vertical grey band.
Indeed, the knee of the luminosity function at
$z=1.5$ (in the middle of the redshift range $z=$1--2) corresponds to
the flux densities at which the flattening in the number counts
occurs. Similarly, the stellar and dust mass at which the flattening
occurs in the number counts corresponds to the knee of the respective
mass functions at $z=1.5$. We furthermore find that the faint--end slope of the
dust continuum luminosity functions (and dust mass function) is significantly shallower than
the low--mass slope of the stellar mass function (almost flat at $z<2$;
compare the top two
panels to the bottom left panel). This is driven by the strong dependence of the gas--phase metallicity on stellar mass and the strong dependence of the dust-to-gas ratio on the gas-phase metallicity.  Because of this shallow slope in the dust continuum luminosity function,
integrating to fainter flux densities results in only a modest increase
in detected sources, as will be discussed in Sec. \ref{sec:discussion}. The flattening in the number counts thus corresponds to probing galaxies below the knee of the luminosity function.

Our model assumes that a set of empirical relations can be used to
describe the entire population of galaxies from low to high
redshifts. It is therefore
worthwhile to explore if our finding that the flattening in the number
counts is caused by the shape of the dust continuum luminosity function is robust against changes in the assumed empirical
relations. In Appendix \ref{sec:assumptions} of this work we adopt a
variety of different assumptions, including different recipes to
assign gas masses to galaxies, different mass-metallicity relations, a
different assumption for the amount of star formation that is dust
obscured, and different assumptions for the dust--to--gas ratio of
galaxies. Every empirical relation used in the model has an error
associated to it. To better understand how the error in these
components affects the number counts we run the model 100 times, sampling over
the intrinsic error for each empirical relation. The different assumptions change the
normalization of the number counts by up to a factor of two. It
furthermore slightly changes the shape of the cumulative number
counts. Nevertheless, for none of the explored
scenarios does the flattening in the number counts disappear. In other
words, this flattening is not driven by changes in the assumptions on how we derive the dust-to-gas ratio of galaxies, their gas mass, the fraction of obscured
star-formation, their metallicity, or the uncertainties in the individual model components. 
This strengthens our conclusion that the flattening in the
number counts is simply caused by the distribution of the underlying galaxy
population, i.e., probing galaxies below the knee of the dust continuum luminosity functions/mass functions.

\begin{figure*}
\centering
\includegraphics[width = \hsize]{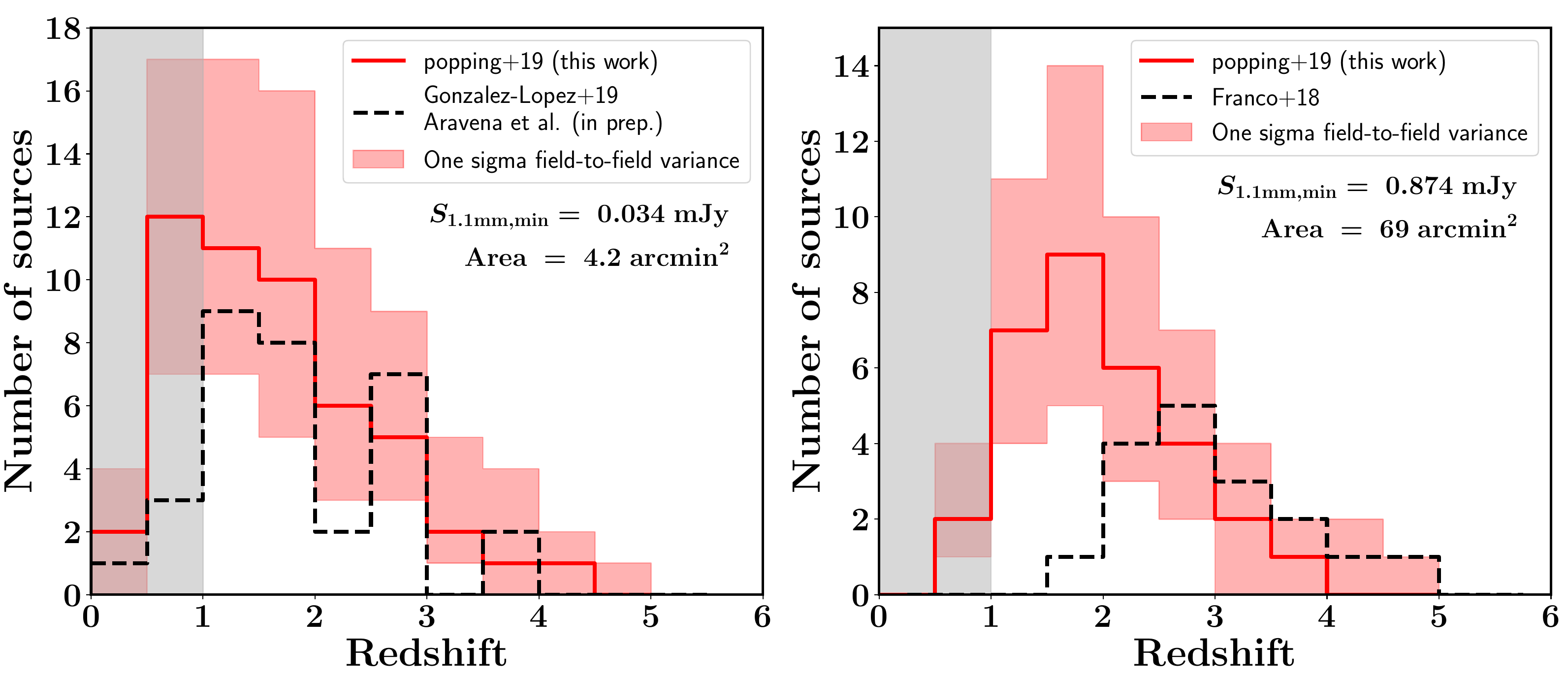}
\caption{A comparison between the predicted and observed redshift
  distribution of galaxies observed at 1.1 millimeter. To account for field--to--field
 variance, we calculate the number counts 1000 times over
 a random portion of the entire modeled lightcone covering the same
 area as the observations, imposing the same survey depth (as outlined
 in the individual panels). The solid line corresponds to the median redshift
  distribution, whereas the shaded region corresponds to the one-sigma
  scatter. Model predictions are compared to the observations by
  \citet[left]{Gonzalez2019numbercounts} (and Aravena et al in prep.) and
  \citet[right]{Franco2018}. The gray shaded area (at $z<1$)
  marks the regime where the model predictions can not be fully
  trusted because the negative k--correction does not apply anymore at
  those redshifts.   \label{fig:redshift_distribution_literature}}
\end{figure*}

\subsection{Redshift distribution}
Current (sub-)millimeter surveys with ALMA have predominantly detected
galaxies at redshifts $z<3.5$ (see for example Figure 18 in
\citealt{Franco2018} and other figures in \citealt{Aravena2016} and \citealt{Bouwens2016} and \citealt{Gonzalez2019numbercounts}).  Even though
ALMA has pushed the detection limit of galaxies to flux densities
below 0.1 mJy, the fraction of galaxies at redshifts larger than 3.5
still remains very low. This is driven by the dominant contribution of
galaxies at $z=1-3$ to the number counts (Fig. \ref{fig:number_counts_redshift}). 

To quantify the agreement
between the redshift distribution of (sub-)mm detections  predicted by
our model and the current observations, we present a comparison
between the two in Figure
\ref{fig:redshift_distribution_literature}. For this comparison, we
adopt the same field--of--view and sensitivity cutoff as the observations. We compare our predictions to observational
results by 
\citet{Franco2018} and \citet{Gonzalez2019numbercounts}. These works
probe the 1.1 millimeter number counts over an area of 69 arcmin$^2$ \citep{Franco2018} down to
 0.874 mJy and an area of 4.2 arcmin$^2$ down to 0.034 mJy
 \citep{Gonzalez2019numbercounts}. To account for field--to--field
 variance, we calculate the number counts  1000 times over
 a random portion of the entire modeled lightcone covering the same
 area as the observations (similar to Figure
 \ref{fig:number_variance}). We show the mean and one-sigma distribution
 of the predicted number counts. The predicted redshift distribution
 at $z<1$ can not fully be trusted, as the negative k--correction
 implied by our model does not apply at these redshifts.

Overall we find that the observed redshift distributions from \citet{Gonzalez2019numbercounts} typically all
fall within the one-sigma scatter of the model predictions. This suggests that, at
least at $z<3$, the model not only successfully reproduces the
cumulative number counts of galaxies, but also the redshifts of the
sources that are responsible for these number counts. The low-number
statistics of detections at $z>4$ makes it hard to further quantify
the success of the presented model. Possibly most surprising is the
lack of sources detected by \citet{Franco2018}  at $z<2$ compared to our model predictions. We additionally  find that at $\sim 1$ mJy, our model predicts number counts higher than derived by \citet{Franco2018}. Given the success of our model in reproducing the number counts by \citet{Gonzalez2019numbercounts}, the apparent mismatch with Franco et al. may suggest a tension between the model predictions and observations for the brightest millimeter sources, but we note that not all sources in the \citealt{Franco2018} sample have a spectroscopic redshift. Furthermore, a prior based selection of the data presented in Franco et al. suggested that additional sources may have been missed in the blind selection, which may change the redshfit distribution (Franco et al. in prep). Lastly, it  has to be noted that the observations still fall within the two--sigma range of the model predictions. Our model predicts a higher median redshift for a survey similar to \citet{Franco2018} than \citet{Gonzalez2019numbercounts} (although the median redshift predicted for a survey with the Franco et al. specifics is different from what was observed). This is in agreement with previous findings that the survey depth can significantly alter the redshift distribution with shallower surveys yielding higher mean redshifts \citep{Bethermin2015}.

\section{Discussion}
\label{sec:discussion}
\subsection{Observational consequences}
\label{sec:discussion_obs_consequences}
We have presented a new data driven model for the cumulative number counts
and redshift distribution of (sub--)millimeter detections of
galaxies. This model successfully reproduces current observations (the cumulative number counts, number counts in bins of different galaxy properties, and redshift distribution functions),
including the flattening in the 1.1 millimeter number counts observed
by \citet{Gonzalez2019numbercounts}. There is a simple origin for this
flattening, namely the shape of the  underlying luminosity function of
galaxies at 1.1 millimeter in the redshift range between $z=1$ and $z=2$
(probing the knee and shallow faint end slope). We
have furthermore demonstrated that this conclusion is
robust against field-to-field variance and the assumptions made in the presented
model. The predicted (and observed) flattening in the number counts has
clear consequences for future continuum surveys with ALMA. A survey at 1.1 millimeter
deeper than 0.1 mJy will not significantly increase the number
of detected sources per square degree.  A similar flattening is to be expected
for the  850 \micron number counts below 1 mJy, a flux
density regime only probed by \citet{Oteo2016} so far. Given our
predictions, a future deep survey at 850 \micron will detect fewer sources
than naively have been expected when extending a simple fit to the
current 850 \micron number count observations.

We can further quantify this by looking into the expected results of hypothetical surveys. In Figure \ref{fig:hypothetical} we show the expected number of sources for a survey covering a given area to a given depth. We furthermore show how many hours per pointing it takes to reach that depth (adopting a signal-to-noise ratio of three and assuming standard ALMA assumptions in the respective bands with 50 antennas), and how many pointings are needed to cover the targeted area adopting Nyquist sampling. On the top two panels, we also plot contours that mark a fixed number of expected detections. As expected, an increase in area and an increase in depth both result in a larger number of detected galaxies. Below 0.1 mJy (for 1.1 millimeter, 0.3 mJy for 850 \micron) the contours of constant number of sources are almost horizontal (i.e., scale less strongly with sensitivity than with area). An increase in the depth from 0.1 to 0.01 mJy  only results in an increase of a factor of $\sim$3 in the detected number of sources. An increase of the area with an order of magnitude naturally results in an increase of a factor 10 in the detected number of sources. This suggest that if the goal of the survey is to detect large number of sources for better statistics, an increase in area is more effective than an increase in survey depth once one has reached a depth of $\sim$0.1 mJy at 1.1 millimeter ($\sim$ 0.3 mJy at 850 \micron).

In the bottom two panels of Figure \ref{fig:hypothetical}, we show contours of fixed total on source time necessary to perform such a survey. This clearly shows that to detect a large number of sources for proper statistics a wide survey is more time efficient than a deep survey. Figure  \ref{fig:hypothetical} also shows that although galaxies are intrinsically brighter at 850 \micron, a survey at 1.1 mm is actually more time efficient. Because the primary beam of ALMA at 1.1 millimeter is larger than at 850 \micron, within a fixed time a survey at 1.1 millimeter can detect fainter sources over a given area than a survey at 850 \micron (as the time is distributed over fewer pointings and thus a fainter sensitivity limit can be reached). The number of expected detected sources per square arc minute is roughly the same between a survey at 850 \micron and 1.1 millimeter for a fixed on source observing time.

In Figure \ref{fig:redshift_distribution} we plot the redshift
distribution of galaxies per arcmin$^2$ for surveys reaching different
depths. We explore the
redshift distribution when accounting for galaxies with flux densities
brighter than 0.01, 0.1, and 1 mJy, respectively. We mark the redshift range $z<1$ with a grey vertical band, as the negative k--correction assumed in our model does not apply for this redshift range.

As the depth of the survey increases, the number of galaxies per
arcmin$^2$ increases at every redshift. The number of galaxies
detected per arcmin$^2$ is systematically higher at 850 \micron than
at 1.1 mm by a factor of three for a survey down to 1 mJy and a factor of
1.5 for a survey down to 0.1 mJy and 0.01 mJy. This is the natural
consequence of the shape of the (sub-)millimeter SED of galaxies,
i.e., lower flux densities at longer wavelengths. Interestingly enough, the
median redshift of the redshift distributions is very similar
for all three survey depths (around $z=1.5$, although note that the uncertain $z<1$ redshift range at which our model may over predict the brightness of sources is included). This seems in tension with observational results (e.g., the higher median redshift of \citet{Franco2018} than \citet{Gonzalez2019numbercounts}), similar to what we saw in Figure \ref{fig:redshift_distribution_literature}.

At 1.1 millimeter, a survey
reaching a depth of 0.1 mJy will detect approximately an order of
magnitude more sources at $1<z<4$ (up to a factor of 30 at
$z\sim5$) than a survey reaching a depth of 1 mJy. An increase in
sensitivity down to 0.01 mJy yields another factor of $\sim3$ increase in
the number of galaxies per arcmin$^2$ at $z>1$.  At 850 \micron a
survey with a depth of 0.1 mJy will detect a factor of 8--10 more
galaxies than a survey with a depth of 1 mJy at $z>1$. An additional
factor of two can be gained by integrating down to a sensitivity of
0.01 mJy. This again emphasises that below flux densities of 0.1 (0.3) mJy at 1.1 millimeter (850 \micron), the number of expected sources only moderately increases with increasing survey depth. At those densities a survey is probing the faint end slope of the dust continuum luminosity function (top two panels Figure \ref{fig:mass_functions}).

Summarising, to significantly increase the number of sources with dust continuum counterparts, a wide survey at 1.1 millimeter at flux density of $\sim 0.1$ mJy is most cost efficient. A gain of only a factor 10 in the number of detected sources compared to the results of \citet{Gonzalez2019numbercounts} could already heavily increase the constraining power for models. Not only will it improve the high-redshift statistics (currently poorly understood), it is also a better approach to obtain dust-continuum counterparts of as many objects as possible that are already detected through optical and near-infrared surveys in common legacy fields. This will allow a more detailed break-down of number counts over different galaxy properties as suggested in this work (e.g., as a function of stellar mass and SFR) and a dust-continuum based gas and dust mass estimate for increasingly large number of galaxies \citep[e.g.,][]{Aravena2016,Scoville2016, Magnelli2019}. The exact survey strategy will ultimately depend on the scientific requirements.


\subsection{What a successful empirical model says about galaxy scaling relations}
Our semi-empirical model combines a data-driven model for the stellar
mass and SFR population of galaxies over cosmic time
\citep{Behroozi2018} with a number of empirical relations to connect
the SFR and stellar mass of galaxies to their dust continuum
emission. It is comforting to realize that this combination correctly
reproduces the observed 1.1 millimeter and 850 \micron number
counts. What this teaches us is that the adopted scaling relations
all seem to hold at least over the redshift regime $z=$0--2
(i.e., the redshift range that most dominantly contributes to the
number counts). This is especially relevant for the adopted relation
between dust--to--gas ratio and gas--phase metallicity and the scaling
between dust mass, SFR, and 1.1 millimeter and 850 
\micron dust continuum flux density, as these
relations have only been observationally probed in this redshift
range for a limited number of massive galaxies \citep[e.g.,][]{Aravena2016,Dunlop2016,Miettinen2017, Aravena2019,Magnelli2019}. We have indeed seen (see Appendix \ref{sec:assumptions}) that a different choice for the
dust--to--gas ratio and mass--metallicity relation results in poorer agreement between the model
predictions and observations. It is furthermore encouraging to see that that the Hayward et al. fitting relations for the dust continuum emission of galaxies result in good agreement with observed number counts, even though these fitting relations were derived for galaxies with flux densities brighter than 0.5 mJy. 

Except for the redshift range between $z=1-2$, the 
constraining power of number counts for our understanding of galaxy physics over cosmic time is rather
limited. The fact that our model successfully
reproduces the redshift distribution of 1.1 millimeter detections up
to $z=4$ (within one sigma) is encouraging, but the low number
statistics in the $z=$2--4 redshift range does not allow us to make
further claims on the validity of the adopted scaling relations in
that redshift regime. It is even harder to make any claims about
the physics at higher redshifts. For example, the contribution of galaxies at $z>4$ to the
number counts is very limited and an order of magnitude increase or decrease in
the number of dusty galaxies at $z>4$ would not change the cumulative number
counts significantly. This suggests that we have almost exhausted what can be learned about galaxy physics from cumulative number counts. It is therefore important that
future observations start to probe the luminosity function of galaxies at discrete
redshifts (and possibly the dust mass function), start connecting the dust continuum  measurement to other galaxy properties, and furthermore aim at resolving the interiors of galaxies at sub-mm wavelengths.  This requires among others complete spectroscopic redshift samples for sizeable numbers of (sub-mm) galaxies. Besides confirming
our theoretical hypothesis about the flattening caused by the knee of
the mass/luminosity functions at $z=$1--2 and the shallow faint end slope, such an effort will provide stringent
constraints currently missing for theoretical models that started to include the
detailed tracking of dust formation and destruction over cosmic time
\citep{McKinnon2017, Popping2017dust, Hou2019, Dave2019}. These
include constraints on the dust mass function, cosmic density of dust,
but also the connection between stellar mass and SFR and dust
properties. An approach to observationally probe the luminosity
function would be to cross-correlate the securely detected dust
continuum sources with information from spectroscopic surveys of the
UDF for example with MUSE \citep{Inami2017, Boogaard2019} or based on
ALMA spectral information \citep{Gonzalez2019}.

\subsection{A top-heavy initial mass function?}
Previous theoretical works have suggested that a top--heavy IMF in starburst environments is necessary to reproduce the number count of bright galaxies, while simultaneously reproducing the optical and near-infrared properties of galaxies \citep[e.g.,][]{Baugh2005,Lacey2016}. Recent observations of active star-forming regions (analogues of high-redshift starbursts) in our Galaxy and the Large Magellanic Cloud \citep{Motte2018,Schneider2018} have suggested that the newly formed stars in these regions indeed have a top-heavy IMF compared to a Chabrier IMF. \citet{Zhang2018} looked at the abundance ratio of isotopologues (an index of the IMF, \citealt{Romano2017}) in $z=2-3$ dust-enshrouded starbursts and concluded that these galaxies have an IMF more top-heavy than a Chabrier IMF. 

We find that we can reproduce the number counts of galaxies at 1.1 millimeter and 850 \micron (up to a few tens of mJy at 850 \micron) under the assumption of a uniform \citet{Chabrier2003} IMF. This is in line with other recent theoretical efforts that suggest that the number counts of sub-millimeter bright galaxies can be reproduced without invoking a top-heavy IMF \citep[e.g.,][]{Safarzadeh2017,Lagos2019}. This does not necessarily mean that starburst environments can not form stars following a different IMF than Chabrier. It suggests that changes in the IMF in order to match sub-mm number counts are degenerate with other ingredients and predictions of galaxy formation models such as the treatment of dust and dust emission and/or the SF properties of galaxies. These degeneracies should be explored with care.

\subsection{Comparison to earlier work}
There have been multiple theoretical efforts in the literature (some of them from first
principles, others adopting a semi-empirical approach similar to our
model) that model the (sub-)mm number counts of galaxies. Pre-ALMA, the focus of these comparisons was on the sub-millimeter galaxies that are orders of magnitude brighter than the sources discussed in this work. Only after ALMA started operations did these comparisons start to include sources with flux densities below 1 mJy. 

\citet{Somerville2012} presented predictions for the 850 \micron
number counts down to 0.01 mJy, based on a semi-analytic model of
galaxy formation \citep{Somerville2008}. This model predicts a sharp drop in the differential
number counts of galaxies for flux densities below 0.1 mJy. The model does not succeed in reproducing the
observational constraints that were available at that time.

\citet{Cowley2016} use a different semi-analytic model to study 850
\micron number counts of galaxies. The authors reproduce the
observations and predict a flattening
in the number counts, but do not explore what causes this
flattening. The authors specifically focus on the effect of
field-to-field variance on observed number counts and similar to us
find that survey design influences how well the underlying `real'
number count distribution of galaxies is recovered.

\citet{Lacey2016} provides predictions for the 850 \micron number
counts using the same semi-analytic model as \citet{Cowley2016}. The
authors specifically explore how different prescriptions for the
baryonic physics in galaxies affect the number counts, but found all explored prescriptions  predict a flattening in the number
counts. This strengthens our conclusion that the flattening is caused by
the underlying galaxy population. The authors furthermore explore the
redshift distribution of sub-mm detected galaxies, but focus on
surveys with a depth of 5 mJy. In order to reproduce the observed
number counts (especially for the brightest flux densities)
\citet{Lacey2016} adopt a top-heavy IMF during starburst events (see
also \citealt{Baugh2005}). Our work on the other hand  suggests that the number counts can be reproduced
by a simple semi-empirical model that does not need to make any
changes to the initial mass function of the stars. 

\citet{Safarzadeh2017} present predictions for the 850 \micron number
counts of galaxies based on a semi-analytic model
\citep{Lu2011,Lu2014}. In this work the authors calculate the 850
\micron flux of galaxies by coupling the SAM output to the fitting
functions presented in \citet{Hayward2013hydro}. The presented
predictions agree fairly well with the observations that were
available at that time (although they seem to predict higher number
densities than found by \citet{Aravena2016} after rescaling to 850
\micron). The model predictions include a flattening of the cumulative
number counts below 850 \micron flux densities $\sim 1$ mJy, in rough
agreement with our predictions. The main result of
\citet{Safarzadeh2017} is that the observed 850 \micron number counts can be reproduced by
the models without invoking the need of a top-heavy IMF, in line with
our findings. This also agrees with the findings using a different semi-analytic model by \citet{Lagos2019}, who reach a similar conclusion by predicting the 850 \micron flux density directly from the star-formation history of the galaxies with a physical model for attenuation.

\citet{Hayward2013hydro} couples a semi-empirical model with the
fitting functions from \citet{Hayward2011} to model the number counts
at 1.1 millimeter at flux densities brighter than 0.5 mJy. The model
reproduced the available constraints at that time, but did not look at
faint enough galaxies to probe the existence of the flattening in the
1.1 millimeter number counts. The authors furthermore present the
redshift distribution function for a survey at 1.1 millimeter with a
flux density sensitivity of 1.5 mJy and find a median redshift of
$z=3$, with a quick drop at $z>4$. This median redshift is higher than
predicted by our model. The origin of this difference may lie in the
adopted approach to estimate the dust mass of
galaxies. \citet{Hayward2013hydro} adopt a fixed dust--to--metal
ratio, a different mass--metallicity relation, and a different
approach to estimate the gas mass of galaxies. As demonstrated in the
Appendix of this paper (see Figure \ref{fig:number_counts_methods}),
these different approaches result in changes in the normalization of
the number counts and small changes in their shape. Especially given
the difference between the \citet{Zahid2013} and \citet{Maiolino2008}
mass--metallicity relation, it is not surprising that this leads to a
different redshift distribution.

Similar to the work presented in this paper, \citet{Hayward2013SHAM}
coupled the fitting functions from  \citet{Hayward2011} to the
sub-halo abundance matching model presented in
\citet{Behroozi2013}. Hayward et al. were particularly interested in
the effects of blending (i.e., spatially and physically unassociated
galaxies blending within one beam) on the derived 850 \micron number
counts of single-dish surveys and found that, indeed, for single dish
surveys blending contributes significantly to the number counts at
flux densities brighter than 2 mJy (the exact contribution of blending
to the bright end of the number counts depends on the adopted beam
size). In this work we are mostly comparing our model predictions to
observations that probe fainter regimes (fainter than 2 mJy at 850
\micron) where blending is less of an issue and/or based on ALMA
results, for which the beam size is sufficiently small to easily separate
the individual sources.

\citet[see also \citealt{Bethermin2012}]{Bethermin2017} developed a
semi-empirical model for the number counts of galaxies. This model is
conceptually similar to the work presented here, but also accounts for
the effect of lensing on the number counts of galaxies. The authors
find a flattening in the 1.2 millimeter number counts at flux
densities below 0.1 mJy, although not as strong as we find and
suggested by observations. The authors furthermore explore the
redshift distribution of galaxies, exploring a scenario with a survey
depth of 4 mJy at 850 \micron and 1.5 mJy at 1.2 millimeter (see also
\citealt{Bethermin2015}). \citet{Bethermin2017} find that for the
latter scenario the redshift distribution peaks at around $z=$2--3,
slightly higher than our findings. The authors do not aim to explore what the
properties are of the galaxies that contribute to the number counts at
different flux densities.

\citet{Casey2018} also presented a model for the (among others) 1.1
millimeter and 850 \micron number counts. Casey et al. explore a
number of star-formation history scenarios (especially focusing on the
fraction of dust-obscured SF at $z>4$) and investigate how these
changes in the star-formation histories manifest themselves in the
(sub--)millimeter number counts. The authors do not focus on flux
densities faint enough to discuss their theoretical predictions for a
flattening in the number counts. 

\section{Conclusions}
\label{sec:conclusion}
In this paper we presented a semi-empirical model for the number
counts of galaxies at 1.1 millimeter and 850 \micron. This model is based upon the \textsc{UniverseMachine} (\citealt{Behroozi2018}, a model that predicts the stellar mass and SFR distribution of galaxies over cosmic time) with theoretical and empirical relations that predict the dust emission of galaxies as a function of their SFR and dust mass. This model can explain the observations at flux levels that were not reachable pre--ALMA. We summarise our
main results below.

\begin{itemize}
\item The predictions by our fiducial model are in good agreement with the observed cumulative number counts and number counts in bins of different galaxy properties. The model reproduces the flattening observed in the 1.1
millimeter number counts of recent deep surveys with ALMA. A similar
flattening is predicted for 850 \micron number counts below 1 mJy.

\item  We demonstrate that the flattening in the 1.1
  millimeter number counts reflects the shape of the underlying
galaxy population at $z=$1--2, i.e., the observations are probing the
knee and the shallow faint end slope of the
1.1 millimeter luminosity function.  

\item The galaxies at the `knee' of the  1.1
  millimeter number counts have redshifts between $z=1$ and $z=2$,
  stellar masses around $5\times 10^{10}\rm{M}_\odot$ and dust masses
  of the order $10^8~\rm{M}_\odot$. 

\item The observed ASPECS redshift distribution of 1.1 millimeter ALMA detections is
  in agreement with the model predictions after we account for field--to--field variance.

\item Future dust continuum surveys at 1.1 millimeter and 850 \micron
  surveys that aim to detect large numbers of sources through their dust emission should cover large areas on
  the sky once below a flux density of $\sim$0.1 mJy (at 1.1 millimeter, $\sim$ 0.3 mJy at 850 \micron), rather than integrating to faint flux densities over small
  portions on the sky. 
  
\item Our model successfully reproduces the number counts of galaxies without the need to adopt an IMF different from \citet{Chabrier2003}. This is in contrast with theoretical models suggesting that a top--heavy IMF is responsible for the observed number counts of bright millimeter galaxies.

\item The success of our model to reproduce the number counts of
  galaxies suggest that the adopted empirical relations in our
  fiducial model (to estimate
  the gas mass, the gas-phase metallicity, obscured fraction of star
  formation, dust mass, and dust continuum flux of galaxies) are valid
  up to $z=2$. Different choices for the empirical relations lead to
  poorer agreement with the observations.

\end{itemize}

The success of our model to describe the number counts of galaxies
at 1.1 millimeter and which galaxies are responsible for these number
counts also means that we have exhausted the amount of information about galaxy physics that can be
extracted from dust continuum number counts. Mainly
because the number counts are biased towards a narrow redshift range
from redshift one to two. To further our knowledge about galaxy physics
from continuum observations, future observational efforts should focus on the dust continuum properties in discrete redshift bins (e.g., dust continuum luminosity function), as a function of other galaxy properties, and on spatially resolved, multi--band dust continuum properties of galaxies and their connection to the resolved stellar and gas properties
of galaxies. 

\acknowledgments
We thank Caitlin Casey, Philipp Lang, Desika Narayanan, and I-Ting Ho for useful discussions and Claudia Lagos and especially Ian Smail for comments on an earlier version of this paper. We additionally thank the referee for constructive comments. Computations for this work were performed on Rusty at the Center for Computational Astrophysics, Flatiron Institute. The Flatiron Institute is supported by the Simons Foundation. F.W. acknowledges support from ERC Advanced Grant 740246 (Cosmic Gas). Este trabajo cont\'o con el apoyo de CONICYT + Programa de Astronom\'ia+ Fondo CHINA-CONICYT CAS16026. Este trabajo cont \'o con el apoyo de CONICYT + PCI + INSTITUTO MAX PLANCK DE ASTRONOMIA MPG190030. R.J.A. was supported by FONDECYT grant number 1191124. F.E.B. acknowledges support from CONICYT-Chile (Basal AFB-170002, FONDO ALMA 31160033, FONDECYT Regular 1190818)), the Ministry of Economy, Development, and Tourism's Millennium Science Initiative through grant IC120009, awarded to The Millennium Institute of Astrophysics, MAS. D.R. acknowledges support from the National Science Foundation under grant number AST-1614213 and from the Alexander von Humboldt Foundation through a Humboldt Research Fellowship for Experienced Researchers. This Paper makes use of the ALMA data \newline ADS/JAO.ALMA\#2016.1.00324.L. ALMA is a partnership of ESO (representing its member states), NSF (USA) and NINS (Japan), together with NRC (Canada), NSC and ASIAA (Taiwan), and KASI (Republic of Korea), in cooperation with the Republic of Chile. The Joint ALMA Observatory is operated by ESO, AUI/NRAO and NAOJ. The National Radio Astronomy Observatory is a facility of the National Science Foundation operated under cooperative agreement by Associated Universities, Inc.

\bibliographystyle{aasjournal.bst}
\bibliography{references}

\appendix
\section{Is the flattening in the number counts robust against the
  assumptions made in the semi-empirical model}
\label{sec:assumptions}
\begin{figure*}
\centering
\includegraphics[width = \hsize]{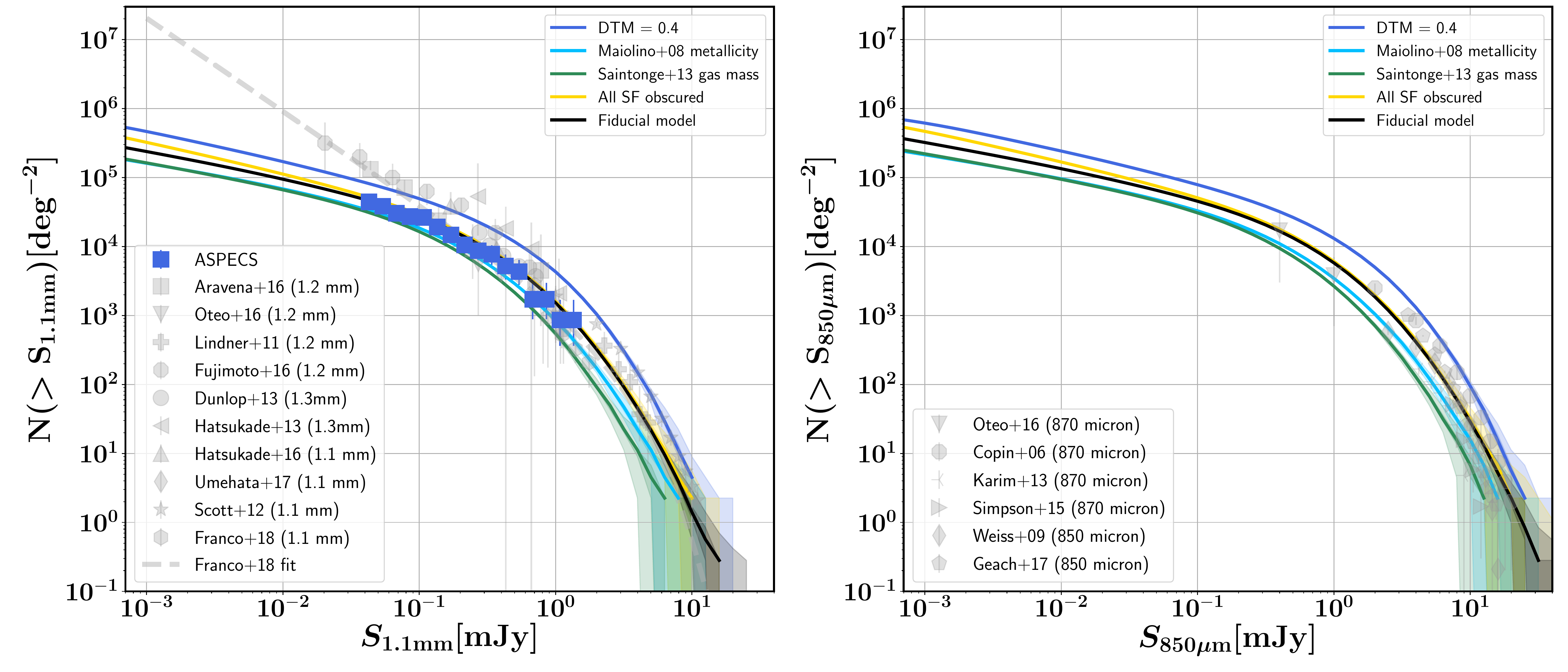}
\caption{The observed and predicted 1.1 millimeter
  and 850 \micron galaxy  number counts. The black solid
  line marks the fiducial model discussed in this paper. The coloured lines mark the number counts when
  replacing individual components of the model by different empirical
  relations/assumptions  discussed in the Appendix. The shaded region marks the one-sigma variance of the 100 random
  realizations when sampling over the error of the individual
  components of the model. There are some changes in the
  normalization of the  number counts when
  varying individual components of the model within a facor of $<$2, but overall the shape of
  the number counts is robust against the changes applied to the model.
\label{fig:number_counts_methods}}
\end{figure*}

In the main body of this paper we have connected the predictions from the
\textsc{UniverseMachine} to a number of empirical relations to
estimate the sub-mm flux density of galaxies. In this Appendix we explore how robust our results are
against the exact choice in these empirical relations. We replace the
empirical relations in our fiducial model by other relations/assumptions proposed in
the literature and show the resulting predicted number counts in
Figure \ref{fig:number_counts_methods}.

\paragraph{Gas masses estimated following \citet{Saintonge2013}}
We have adopted the methodology presented in \citet{Popping2015SHAM}
to estimate the gas mass (atomic plus molecular) of galaxies. An
alternative option is the fit for the \h2 mass of galaxies as a
function of stellar mass, SFR, and redshift
given in (\citealt{Saintonge2013}, note that this prescription does
not include a contribution by \hi to the total gas mass). We find that
the number counts are systematically a factor 1.5--2 below the
predictions of our fiducial model.
 
\paragraph{Fixed dust--to--metal ratio of 0.4}
Theoretical models typically make the assumption that the
dust-to-metal ratio of the ISM equals 0.4. When adopting the same value
(thus not scaling the dust-to-metal ratio of the ISM as a function of
the gas-phase metallicity) the predicted number counts are a factor of
1.5--2 above the predictions by our fiducial model. Although the
overall normalization of the number counts changes, the flattening does
not disappear.

\paragraph{Mass-metallicity relation from \citet{Maiolino2008}}
An alternative fit of the gas-phase metallicity of galaxies as a
function of their stellar mass and redshift in the redshift range from
$z=0$ to $z=3.5$ was presented in
\citet{Maiolino2008}. We adopted the \citet{Zahid2013} relation for
our fiducial model as this is based on a more robust sample of galaxies with a coherent
metallicity calibration. The number counts predicted when adopting the
\citet{Maiolino2008} mass-metallicity relation are a factor $\sim1.5$
below the predictions by our fiducial model.

\paragraph{All star-formation is obscured}
We adopted the fit presented in \citet{Whitaker2017} to estimate the
obscured fraction of SF. An extreme alternative is to assume that all
SF happens in dust environments and $f_{\rm obscured} = 1$. We find
that the resulting number counts are essentially the same as predicted
by our fiducial model, expect for the faintest flux densities. 

Summarising, we find that the exact choice for the individual
components of our model change the normalization of the number counts,
but not the presence of a flattening. This confirms that the flattening
seen in the data is indeed a result of the underlying galaxy
population and not due to the adopted approach to assign sub-mm luminosities to galaxies.

\section{A hypothetical survey}
\begin{figure*}
\centering
\includegraphics[width = \hsize]{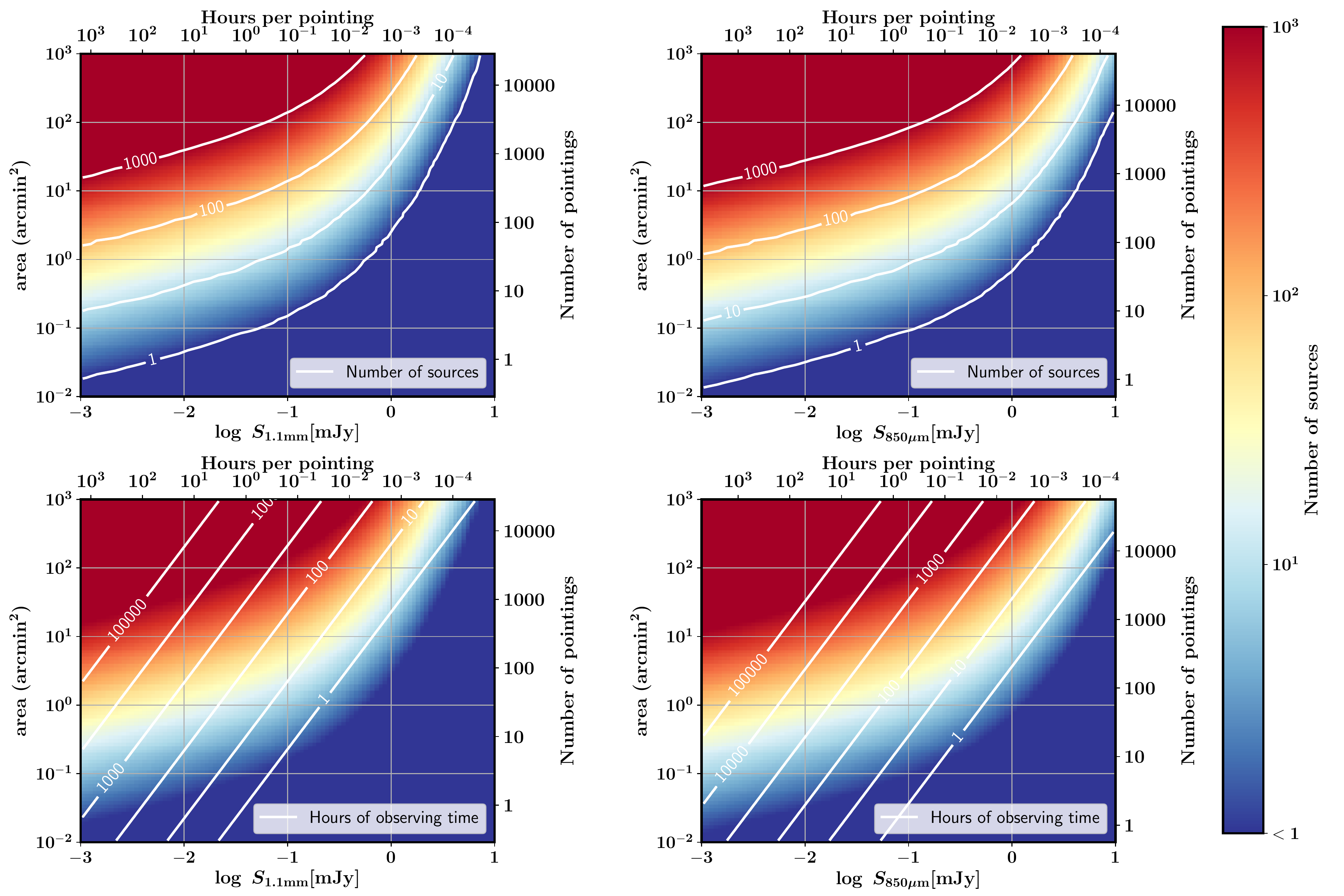}
\caption{The expected number of sources for a hypothetical survey at 1.1 millimeter (left column) and 850 \micron (right column), as a function of the survey depth and covered area, as well as the number of hours per pointing it takes to reach this depth (at a signal-to-noise ratio of 3) and the number of pointings necessary to cover the area assuming Nyquist sampling (all assuming standard ALMA assumptions for 50 antenna's). In the top row, contours depict lines of a fixed number of expected sources. In the bottom row, contours depict a fixed total on source observing time. Below flux densities of 0.1 (0.3 mJy) a wide survey at 1.1 millimeter (850 \micron) is more (cost-)efficient to increase the number of detected source than a deep pointed survey.
\label{fig:hypothetical}}
\end{figure*}

\begin{figure*}
\centering
\includegraphics[width = \hsize]{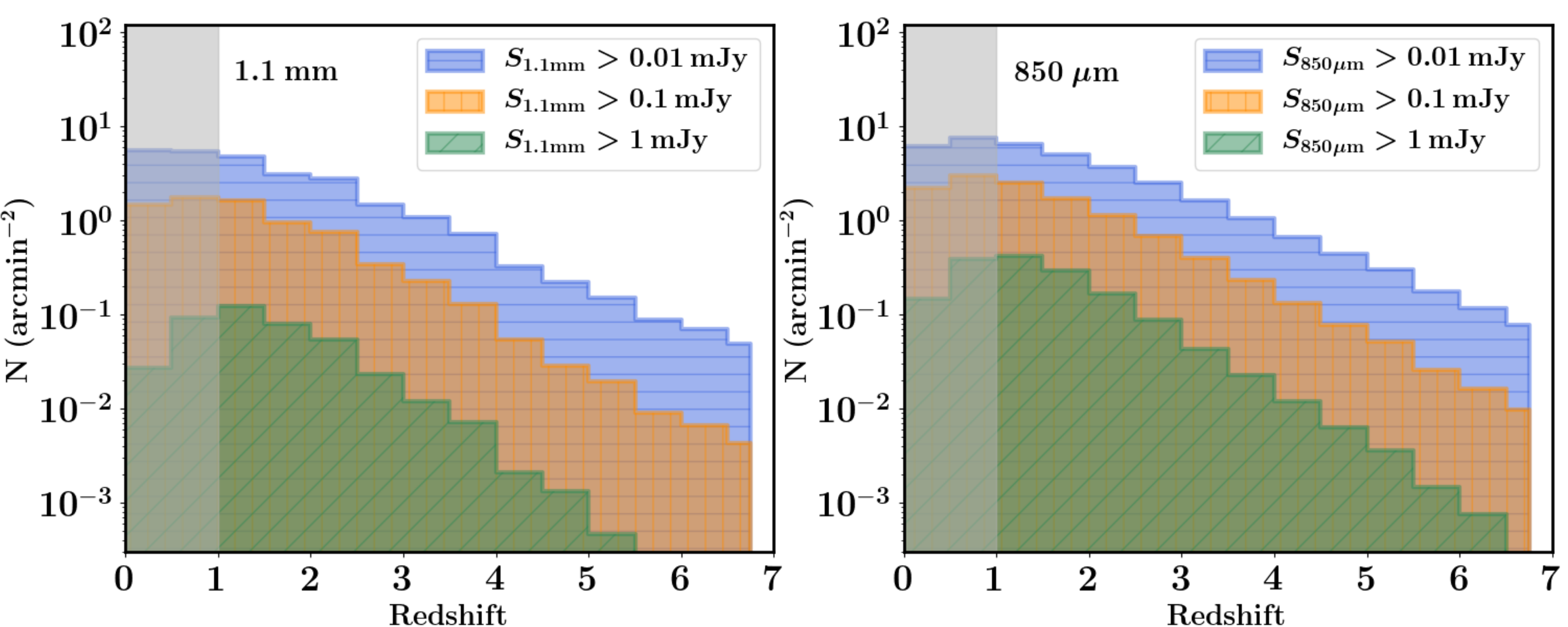}
\caption{The redshift distribution of galaxies as a function of their
  1.1 millimeter  (left) and 850 \micron (right) flux density.  A different survey depth results in preferably detecting
  galaxies at different redshift. To efficiently detect galaxies, a
  shallow but wide survey is more time efficient that a narrow but deeper survey. The gray shaded area (at $z<1$)
  marks the regime where the model predictions can not be fully
  trusted because the negative k--correction does not apply anymore at
  those redshifts. 
\label{fig:redshift_distribution}}
\end{figure*}

In Figures \ref{fig:hypothetical} and \ref{fig:redshift_distribution} we show predicted number of observed galaxies and their redshift distribution, respectively, of hypothetical future surveys (with ALMA). These are discussed in detail in Section \ref{sec:discussion_obs_consequences}

\end{document}